# Layers, Folds, and Semi-Neuronal Information Processing

Bradly Alicea[1,2], Jesse Parent[1]

**Keywords:** Semi-cognitive Models, Cognitive Systems, Multiscale Models, Physiological Information Processing

## Abstract

What role does phenotypic complexity play in the systems-level function of an embodied agent? The organismal phenotype is a topologically complex structure that interacts with a genotype, developmental physics, and an informational environment. Using this observation as inspiration, we utilize a type of embodied agent that exhibits layered representational capacity: meta-brain models. Meta-brains are used to demonstrate how phenotypes process information and exhibit self-regulation from development to maturity. We focus on two candidate structures that potentially explain this capacity: folding and layering. As layering and folding can be observed in a host of biological contexts, they form the basis for our representational investigations. First, an innate starting point (genomic encoding) is described. The generative output of this encoding is a differentiation tree, which results in a layered phenotypic representation. Then we specify a formal meta-brain model of the gut, which exhibits folding and layering in development in addition to different degrees of representation of processed information. This organ topology is retained in maturity, with the potential for additional folding and representational drift in response to inflammation. Next, we consider topological remapping using the developmental Braitenberg Vehicle (dBV) as a toy model. During topological remapping, it is shown that folding of a layered neural network can introduce a number of distortions to the original model, some with functional implications. The paper concludes with a discussion on how the meta-brains method can assist us in the investigation of enactivism, holism, and cognitive processing in the context of biological simulation.

## Introduction

The emergence of cognition via the transformations of biological development is characterized by the formation of highly structured neural circuits. As cognition is embodied in the organism, the complexity of its substrate emerges through layering and folding processes, and leads to regulatory processes deriving from this structure. Aside from their role in enabling cognition, layering and folding can be observed in a variety of biological processes across a wide range of organismal systems. This demonstrates the role of these processes in a wider class of intelligent, self-organizing phenomena, which may allow us to propose a universal class of cognitive models. Examples of four types of layering and folding each will be presented, with further consideration of their potential function. Layering occurs in contexts such as the animal embryo, skin, bone, and brain anatomy. Folding, meanwhile, occurs in the form of skin morphogenesis, cortical sheets, rugae, and embryonic expansion. In some cases, such as the Mammalian neocortex, both layering and folding play an important role in the structure and function of the mature organ system.

[1] Orthogonal Research and Education Laboratory, Champaign-Urbana, IL.  bradly.alicea@outlook.com
[2] OpenWorm Foundation, Boston, MA.

The potential to leverage such processes as a means to enrich computational agency will be discussed. In the context of the nervous system [1], it is argued that the neuron doctrine has obscured the role of other types of important functional organization (such as tissue layers). While layering and folding are well-characterized as structural phenomena, the functional significance of each is not particularly well understood. Layering plays a special role in nervous system development, and has been well-studied in this context. In an evolutionary context, layering increases with brain size [2]. Neural layers can also operate semi-independently of one another, allowing for functional robustness [3, 4]. Having multiple layers with functional heterogeneity can also allow for robust control even if the individual layers exhibit suboptimality [4, 5]. In biological and behavioral terms, layered architectures can regulate a variety of functions, including sleep and arousal [6], speed-accuracy tradeoffs [5], and hierarchical mental processes [7]. Meta-brain models [8] offer not only a means to better understand this functional significance, but also to leverage these types of morphogenetic structures for agentive learning and adaptive behavior.

**Layering/Folding as a Composite Process**

In development, it is quite common for both layering and folding to shape morphogenesis of the same tissue. This is understandable, since they result from different causes and do not require an evolutionary nor developmental tradeoff. Layering results from the need for homeostatic control and an optimization functional specialization, while folding results from the need to conserve space within the body plan and conservation of biophysical parameters. In terms of informational regulation between layers, Wilson and Prescott [9] argue that a mechanism called constraint closure provides a means for layered structures in the brain to become functionally integrated. Constraints modify a process, and act to limit that process to a smaller state space [10]. Meanwhile, the dynamics of faster processes act as a scaffold for slower processes.

Overall, layered and folded meta-brain topologies offer some unique topological properties. Our approach is unique with respect to previous studies in that we focus on the representational quality of each layer. We will also focus on how agents can toggle between layers over time, particularly developmental time, which draws parallels with the biological context of [1, 9]. Finally, we will include folding into our functional model, as the combination of folding and layering provides an added dimensionality to layered control architectures proposed for embodied agents [4]. These issues will be explored using a meta-brain model of the Mammalian gut. It can be argued that this type of semi-neuronal system can be understood in terms of representationalism.

**Layering**

Layering (or laminar organization) is a hallmark of organismal development. From the first cell types of the embryo, the resulting tissue sort themselves into layers rather than remaining a mosaic structure. Laminar organization allows for selective functional organization, and occurs in a variety of tissue types and organs.

**Germ Layering.** One example of layering comes from the germ layering that occurs during early embryogenesis. In animal embryos, the blastocyst goes on to form either a two- or three-layered germinal structure. Each of these germ layers go on to form a set of differentiated tissues that do not resemble the layering of early development. Yet the layered origins of these tissue types allow for a multitude of tissue types and asymmetric configurations from this root model of differentiation.

**Topographical Anisotropy of Skin.** Another example of layering, this time in a differentiated organ, is skin. In human skin, there are three layers: stratum, papillary, and reticular. Layers of skin provide a means for structural integrity, self-repair, and as a barrier against pathogens and infection. These layers are organized topographically (top to bottom), and have their origins in the ectoderm. Cells originating in this layer of the embryo undergo a process of stratification, which is also important for maintaining local homeostasis [11].

**Topographical Anisotropy of Bone.** Skeletal bone also provides an example of layering. Skeletal bone typically has four layers: periosteum, cancellous, marrow, and cortical. Developmentally, a layered structure allows for cartilaginous precursors to be replaced with more derived forms, thus suggesting a role for developmental constraints. As with skin, the layering of bone provides mechanisms for homeostatic maintenance and structural stability. In the case of Mammalian limbs, laminar bone also provides an anisotropic substrate for muscle function and movement.

**Laminar Organization of the Brain.** Our final example of layering is probably the most expansive. In brain anatomy, it is observed that several structures have variable layers. Human neocortex has six layers that have distinct roles in a vertical processing stream. These functional streams form columns that can serve as information processing units [12, 13]. Grossberg [14] refers to this arrangement as laminar computing, or the layered organization of top-down and bottom-up processes. This arrangement is found in a number of other Mammalian brain structures, including the cerebellum, the retina, and the brainstem. As with the neocortical example, these layers typically consist of cell populations that relay information to other regions. This differs from other types of layering in that the main function of anatomical layering in the brain is to process information.

**Folding**

Folds occur as a consequence of buckling or some other spatial displacement. The folding process often occurs as a consequence of growth, development, or disease. Folding results from physical processes such as growth or change in the surface properties of the tissue (substrate). Active folding often occurs in conjunction with layering in a variety of tissue types and organs.

**Embryogenetic Folding.** In embryogenesis, folds occur as a consequence of growth. While tissues can expand as they grow, they are often constrained by the compactness of the egg. Examples of folding in during embryogenesis include formation of the neural tube and gut invagination/evagination. These folding processes help to govern shape regulation during cell proliferation [15]. More generally, folding helps to coordinate a rapidly expanding physical system. When embryos are perturbed during convergent extension (part of a folding process), axial deformations due to stress and strain result [16].

**Neural Convolutions, Rugae, and Skin Folds.** Cortical sheets, which are subject to neural convolutions, also have a propensity to fold. Some of this is due to the nature of folding as a phylogenetic trait [17], but can also be accentuated due to rapid brain growth [18]. Interestingly, folding seems to be constrained by the relative thickness of the layered substrate rather than solely by cell number expansion [19]. Rugae, or folds along the edges of a tubular structure, can be found in anatomical locations as diverse as the stomach, intestine, and vaginal canal. Rugae enable these structures to increase their surface area while equalizing internal pressure, all the while not exceeding anatomical size constraints of an organism's body size. This is similar to cortical folding, where the neocortical sheet must fold so as to not exceed the size limitations of the skull. In the stomach and duodenum, the material properties of the substrate (relative elasticity) affect the ability of the tissue to stretch. Increased rugosity (folding) helps with stretching during digestion [20]. Skin folds are formed in development by arbitrary forces. As in the case of convoluted cortical sheets, the gyri and sulci can result purely from physical processes [21]. Similar to the functional role of rugae, folds provide enough slack for a relatively elastic tissue to stretch. In the case of skin folds, this tissue must stretch around a joint at large angles of displacement without tearing.

**Meta-brains and Semi-neuronal Representations**

Meta-brain models are embodied hybrid models made up of computational components with different degrees of representation [8]. These components can be formal models such as connectionist networks, or symbolic models such as ontologies. This particular type of composition is similar to so-called neuro-symbolic models [22]. What is unique about meta-brain models is the ability of representation-rich layers to encase or have an explicit topological correspondence with representation-free layers. The encasement of meta-brain layers is further embodied through a sparse set of communication pathways that emerge via the principles of differential growth and activity-dependent plasticity, even though the layer is composed of a formalized computational model. In a biological system, evolutionary change proceeds by adding new components, which can be cells [23] or layers of a ganglion. In fact, there is a phylogenetic relationship between new layers and increased neural complexity [24]. Much of this complexity is due to the need to find new or parallel pathways for regulating behaviors [9, 25]. These biological principles are utilized in a meta-brain to add topographical organization to an artificial system.

**Biological Information Processing**

One example of a semi-neuronal representation is the gut (Figure 1). The gut is often referred to as a mini-brain [26], as it forms the core of the enteric nervous system and exhibits both layering and folding. Not only does the gut exhibit mature anatomical folding in the form of rugae, but also a product of invagination during gastrulation. While there are different functional layers of cells, these layers are a mix of neuronal and non-neuronal cells. While the neuronal cells are specialized for the transmission of information, non-neuronal cells exhibit movement and significant intercellular chemical signaling.

At least three different layers of representation can be defined to represent the folded gut: a spatial map that forms a gradient across the topography of cells, the sensation of mechanical and chemical stimuli, and a local feedback signal that keeps track of the location and intensity of muscle contractions. Each type of representation is then ordered in terms of its representational complexity. The topographic gradient is the lowest level of representational complexity, followed by the presence and absence of local sensation, and then finally a model of activity and feedback (Figure 1).

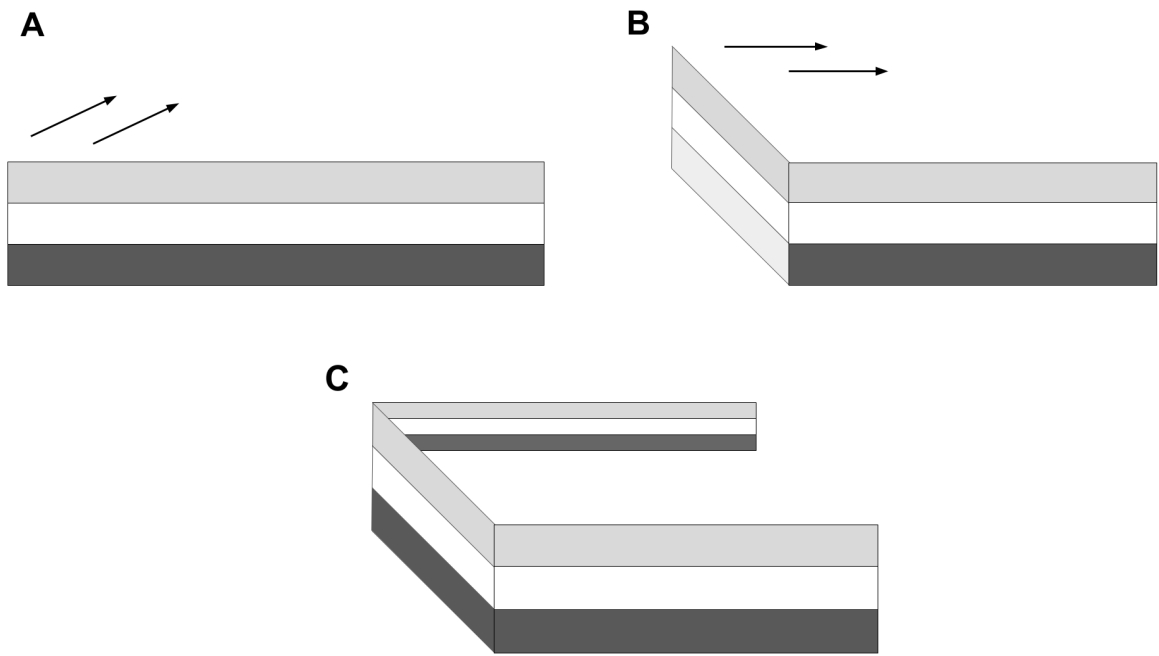

Figure 1. A meta-brain configured as the different anatomically functional and information processing layers of the gut. A: an example of a three layer and unfolded meta-brain. B: a three layer meta-brain undergoing evagination. C: an evaginated three layer meta-brain with an extra fold over the newly-formed gut. Arrows point to the direction of growth and folding.

Figure 1 also demonstrates how a meta-brain can represent developmental morphogenesis as a set of layers. The representational layers are actually quite heterogeneous, but represent a series of precursor and mature tissue layers. One feature of meta-brains is the linkages between layers, which serves to summarize communication and reaction boundaries [27].

**Sub meta-brains: genomic encoding**
A strength of the meta-brain approach is the hybrid nature of different representational levels. Aside from operating as a mechanism that enables neuro-symbolic processing and functional regulation, meta-brains are roughly equivalent to the levels of biological organization [28] defining the basic relationship between genes and phenotype. In meta-brains, there is a sub-representation layer that defines innate aspects of the agent. We use a genomic encoding to reproduce these innate features. Encodings different from representations in that encodings act as a store of compressed information, while representations are a reduced map of some actual phenomenon [29]. Genomic encodings also allow for agent features to evolve given a defined mutation and recombination rate. This can be useful in calibrating these models in the face of representational drift [30], as well as for information and concepts agents have not previously encountered. As a mechanism for expressing compressed information at the right times during development and function, meta-brains utilize a canalization function [31, 32]. Acting as a binary switch, canalization functions respond to ambiguities in the environment in a way that enforces correct and consistent agent behavior. The canalization process restricts us to certain parts of developmental possibility space [33, 34] given the folding and layering trajectory of a meta-brain agent's particular set of representational layers.

Each gene contains information about the shape, deformation, and representational structure, and are expressed with respect to the current state. A canalization-controlled genomic encoding consists of binary sequences (instructions) for each body part, in addition to switches that enable cell proliferation and differentiation, which in turn form the structure of the layers. Additional representation information also exists in this encoding, which is turned on and/or modified throughout the developmental process. Differentiation trees [35, 36] allow us to transform this genetic encoding into a laminar organization. In the example shown for Figure 2, the differentiation tree is used to take a phylogeny of development perspective on agent development. The process of differentiation embodied in the tree structure allows us to specify the sequential specification of an embodied agent and its nervous system. The instructions for this generativity are stored in the genomic encoding.

**Identity of layers in a meta-brain of the gut**

Now we will consider what constitutes a set of biological representations. The lowest level of representation involves simple models of muscle cells, which can be abstracted to force-generated muscle fibers and a source of energy production. This is roughly of the same representationally complex as a connectionist network, and can be implemented using a set of registers and logic gates.

The next most representationally complex layer (above the layer of muscle cells) is a neural network embedding representing sensory neurons and their connections to the central nervous system. An embedding is used for two reasons: the sensory receptors of the gut require spatial heterogeneity and process information in a non-uniform manner. Secondarily, sensory receptor arrays in the gut require the computation of multisensory (pain and chemical) stimuli as well as displacement information. Displacement information is important in muscular hydrostats [37, 38], where muscle ganglia are utilized to manipulate loads. Digested contents must be moved in and out of the gut, requiring sensory communication between the gut surface and a signal that initiates muscle activation.

In a semi-neuronal meta-brain, the processing of multisensory and displacement information from sensation to next available action is accomplished through an implicit feedback mechanism. As part of this mechanism, the highest representational layer draws from the anticipatory biology of Rosen [39]. This layer consists of a predictive model that anticipates muscle activation given the current state and location of material being transported through the gut. This highest layer then communicates with the lowest layer in columnar fashion to update the muscle activational state. These details are summarized graphically in Figure 3.

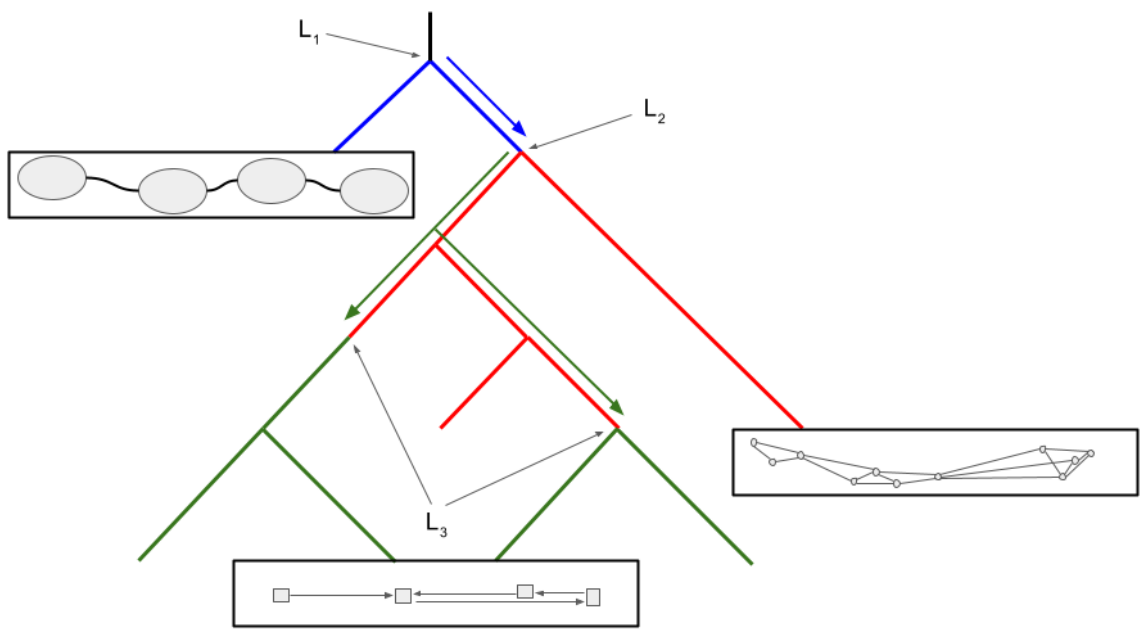

Figure 2. A differentiation tree that describes the morphogenesis of an embodied agent with a layered neural representation. Our tree results from a canalization function. Tree shows progression of development from a single precursor (labeled in black). Branch points ($L_1$, $L_2$, $L_3$) are labeled to show where layers emerge in development. Arrows show a path to the next higher representation level. Branch and arrow colors: Black: precursor. Blue: lowest representational layer (muscle), Red: intermediate representational layer (neural network embedding), Green: highest representational layer (allostatic regulation).

**Local connectivity between layers.** Defining information flow via connections between layers is key to building a topologically-salient representational structure. In this sense, the connections between layers resemble the columnar organization of neocortex [40, 41]. In a meta-brain representation of the gut, columns are defined by specific muscle fibers being connected to local sensory cells for which their sensation is restricted to a specific spatial location on the gut surface. This enables topologically-specific control, and is an important part of capturing a dynamic process of changing shape and functional configuration over developmental time.

**Microbiota and other model interactions.** There are many factors that influence long-term gut function and health that should be included in a meta-brain model, but are not captured by the proposed model. For example, inflammation is a chronic issue that influences the compliance of the gut surface, and sometimes leads to additional folding and/or dysregulation. Microbiota [42,

43] are often used as a biological indicator of these types of interactions. The regulatory role of microbiota might be incorporated into a fourth (and ultimate) representation-rich layer that encodes a computational model of allostatic regulation [44]. Such a model would allow a system-wide indicator of global state, or spatially-distinct indicators of allostatic load. This might be useful in understanding the conditions under which gut dysfunction occurs.

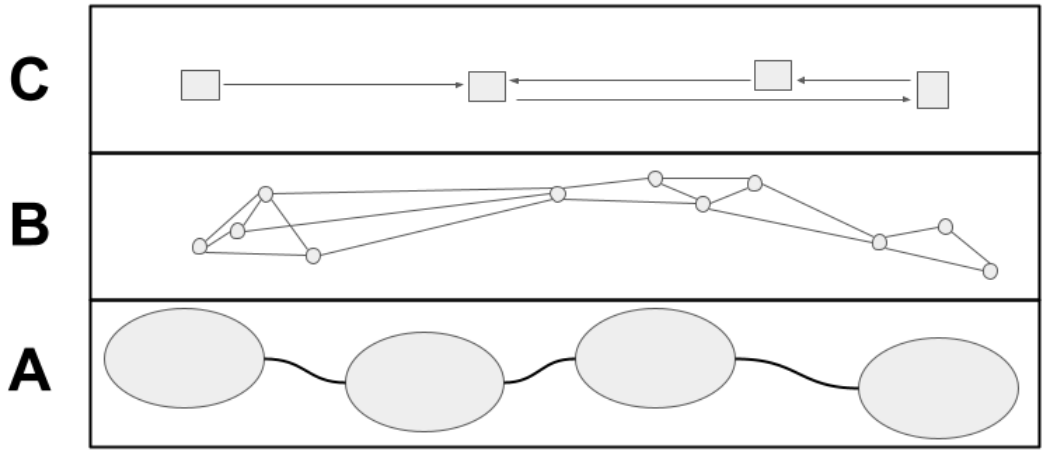

Figure 3. Instantiation of a three-layered gut semi-neuronal meta-brain. A: lowest representational layer (muscle), B: intermediate representational layer (neural network embedding), C: highest representational layer (allostatic regulation).

**Development and morphogenesis of a semi-neuronal representation.** Now that we have defined what a static, adult meta-brain looks like, it is important to consider what the developmental state might look like. This is important when considering how folding affects the initial topographic organization of the layered representation. A developmental trajectory is also important for establishing columnar relationships, or other potential connections between representational layers. The intriguing aspect of this is in modeling the developmental formation of representational models. This might come in the form of adding components or changing the representational contents of each layer. For example, an early developmental meta-brain of the gut might contain muscle fibers with no connections to the sensory layer, while the sensory layer would consist of units that are either disconnected or not indicative of any embedded information (no distinct pattern of connectivity). By contrast, the highest layer (allostatic regulation) may still be in place, but will serve as a summary of behavioral control across the developmental process [45] rather than gut function in a mature model.

**Topological Remapping**
In [8], an example of agent morphogenesis is provided as an example of anatomical fidelity. Another example of anatomical fidelity is the topological convolution that results from the folding of our layered model. One way to demonstrate this is to utilize developmental Braitenberg Vehicles (dBV - [46]) to show how the neural network can fold with growth. We start with a fold wavelength equal to the maximum neural network case size. As the neural network grows in the number of units, the surface of the case size exhibits folding. During this process, the fold wavelength [47, 48] decreases towards a small number, representing a highly convoluted surface. In a layered representational system, these folds introduce distortions that affect both the shortest distance between units and the alignment of processing columns. Figure 4 shows this process of folding and its effects on the alignment of internal representations in the dBV.

**Discussion**

By introducing the concepts of layering and folding to the world of embodied agents, it is demonstrated that these features play important functional, structural and integrative roles in a wide range of biological systems as well as being recapitulated in development. In terms of their utility in artificial life, we demonstrate how a semi-neuronal system (Mammalian gut) can be understood using the meta-brain approach. Multiple types of representation are considered, from simple models of function to predictive and regulative mechanisms. It is further demonstrated that taking a developmental perspective on layering and folding can provide insight into how these processes influence and ultimately change the dynamics of layered and folded systems.

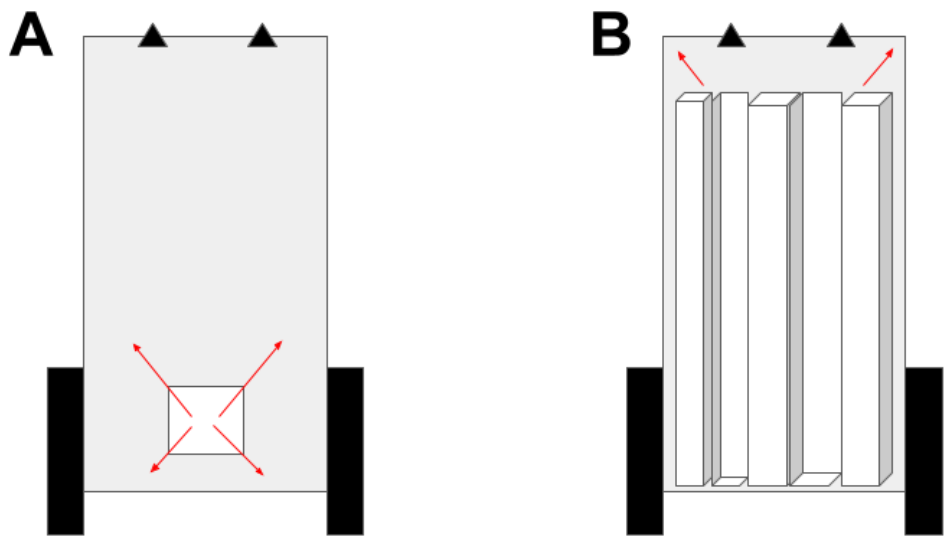

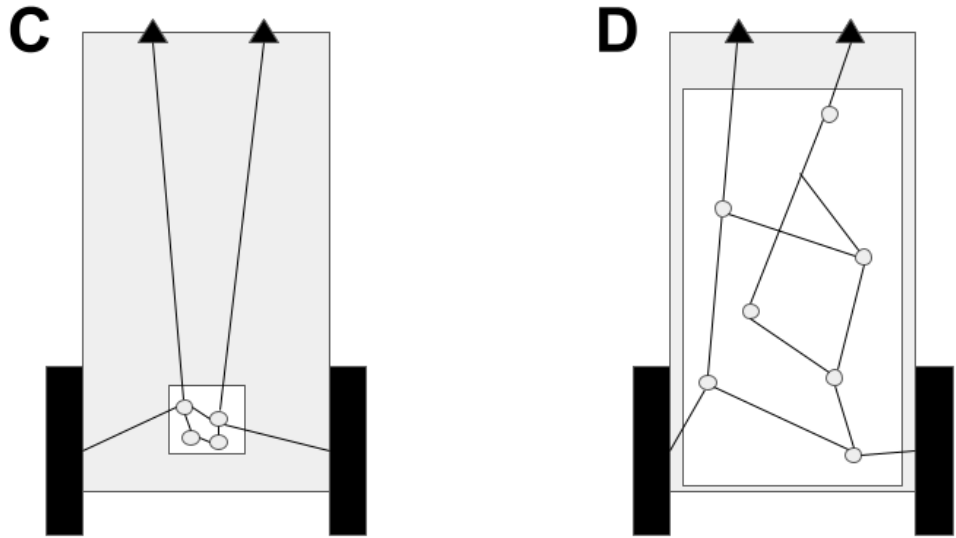

Figure 4. An example of topological remapping as a consequence of folding in a dBV model with a layered neural network. A: a small braincase that expands in size in the direction of the red arrows. B: an expanded braincase that has undergone significant folding as it has expanded and continues to expand (red arrows). C: a single-layered neural network in a small braincase. D: a single-layered neural that has expanded and added neurons. Folding topology is not shown, but reveals displacement of the original nodes and distortion of connectivity.

Applying meta-brains to physiological systems allows for a route to artificial life. Completing this route forces us to think about myriad relational, representational, and organizational issues. In modeling biological systems, meta-brain models require a mechanism for what Heylighen calls relational agency [49]. In a semi- or non-neuronal [50] context, relational agency is summarized in the form of conditional rules and reaction networks. Meta-brain models can use rule systems to communicate between representational layers, and in contexts that are non-neuronal (parts of layers where there is no nervous system communication), such rule systems may be useful.

What does representationalism look like when the things being connected are not neurons engaged in typical cognitive activities? Minimal models [51] often give us a heuristic for thinking about much more complex systems. In the context of Artificial Intelligence, a distinction is made between symbolic and subsymbolic representations. In the case of our gut representation, it is hard to see how symbolic representations are biologically plausible. Yet we can understand systems that manage feedback as symbolic, particularly as it relates to comparing the current state to a historical setpoint. Representations that are allostatic, or that experience significant allostatic load [52], can often exhibit symbolic features.

Meta-brains are called brains because they utilize different layers of mental or symbolic representation. This may seem to be somewhat inappropriate for modeling and simulating biology, as the symbolic continuum does not initially provide much explanatory power. However, representationalism is still an active issue for artificial life systems. According to Halpin [53], biological representations are physically-instantiated structures that create local events and correspond with nonlocal parts of the world. In meta-brain models, this would be implemented as increasingly higher-order transformations of intra-layer processes. Investigating representationalism in a developmentally salient semi-neuronal system opens the door to investigating the role of enaction [54, 55] in the evolution of development.

We conclude by providing two key design principles for semi-neuronal meta-brains. The first is organizational: living systems do not exist as an undifferentiated mass of DNA, proteins, and protoplasm [56]. Biological systems rely upon structural and functional differentiation, and layering is critical to representing this complexity. Yet synthesis (or holism) between multiple parts is also important, and allows for linkages between structures such as layers and folds. Therefore, a second design principle involves establishing strategic connections between different spatial components in biological systems that can be accentuated through layered and folded representations.

## Acknowledgements

Thanks to the workshop and virtual conference circuit for allowing a space to work out these ideas. Thanks also go to Roberto Toro, who suggested the idea of folded neural networks in Braitenberg Vehicles. Additional thanks to the Saturday Morning NeuroSim attendees for additional feedback.